\documentclass[twocolumn,showpacs,nofootinbib,preprintnumbers,amsmath,amssymb,floatfix]{revtex4}
\usepackage{graphicx}
\usepackage{dcolumn}
\usepackage{bm}
\usepackage{longtable}
\usepackage{subfigure}

\begin{document}

\title{Age constraints and fine tuning in variable-mass particle models}

\author{Urbano Fran\c{c}a} 
\email{urbano@ift.unesp.br}

\author{Rogerio Rosenfeld}
\email{rosenfel@ift.unesp.br}

\affiliation{ 
Instituto de F\'{\i}sica Te\'orica - UNESP, Rua Pamplona, 145, 01405-900, S\~{a}o Paulo, SP, Brazil} 

\begin{abstract}
VAMP (variable-mass particle) scenarios, in which the mass of the cold dark matter particles is a
function of the scalar field responsible for the present acceleration of the Universe, have been proposed
as a solution to the cosmic coincidence problem, since in the attractor regime both dark energy and
dark matter scale in the same way.  We find that
only a narrow region in parameter space
leads to models with viable values for the Hubble constant and dark energy density today.
In the allowed region, the dark energy density starts to dominate around the
present epoch and 
consequently such models cannot solve the coincidence
problem.  We show that the age of the Universe in this scenario 
is considerably higher than the age for noncoupled dark energy models, and 
conclude that  more precise independent measurements of the age of the Universe 
would be useful in distinguishing between coupled and noncoupled dark 
energy models.  

\end{abstract}


\preprint{IFT-P.037/2003}

\maketitle

 




\section{Introduction}

The Wilkinson Microwave Anisotropy Probe (WMAP) satellite 
\cite{wmap} has confirmed that the Universe is very nearly flat, 
and that there is some form of
dark energy (DE) that is the current dominant energy component, 
accounting for approximately 70$\%$ of the critical density. DE is
smoothly distributed throughout the Universe and 
is causing its present acceleration \cite{darkenergy,darksector}.
It is generally modeled using a scalar
field, the so-called quintessence models, either slowly rolling 
towards the minimum of the potential
or already trapped in this minimum
\cite{darkenergy,quintessence,plquintessence,expquintessence}. 
WMAP has also given new observational evidence that 
nonbaryonic cold dark matter (DM), a fluid with vanishing
pressure, contributes around 25$\%$ to the
total energy density of the Universe, in agreement with a set of
cosmological observations \cite{darksector,darkmatter}. 

An intriguing possibility is that DM particles could interact with the DE
field, resulting in a time-dependent mass. In this scenario, 
dubbed VAMPs (variable-mass particles) \cite{carrolvamps}, the mass of 
the DM particles evolves according to some function 
of the dark energy field $\phi$, such as, for example, a linear function of the field 
\cite{carrolvamps,quirosvamps,peeblesvamps,hoffmanvamps} with 
an inverse power law dark energy potential or
an exponential function 
\cite{exp1amendola, exp2amendola, exp3amendola,pietronivamps,riottovamps}
with an exponential dark energy potential. 
In this work we are particularly interested in the
exponential case, since it presents a tracker
solution;\footnote{ We adopt the nomenclature ``scaling" solution to denote solutions
where $\rho_\phi \propto a^{-m}$ and $\rho_m \propto a^{-n}$, whereas a
``tracker" solution has $m=n$.} that is, there is a stable attractor regime
where the effective equation of state of DE mimics the 
effective equation of state of DM \cite{exp1amendola,riottovamps}.

This type of solution also appears when the DE  
field with an exponential potential is not coupled to the other 
fluids \cite{expquintessence,exponentialscaling}. 
In fact, Liddle and Scherrer 
showed that for a noncoupled DE, the exponential potential is the only
one that presents stable tracker solutions \cite{tracker}. 
In this case, however, it is 
not able to explain the current acceleration of the Universe and other observational constraints,
unless we assume that the field has not yet reached the 
fixed point regime \cite{jhep}. 

The tracker behavior is interesting because once the attractor
is reached, the ratio between DM energy density $\rho_{\chi}$ and 
DE energy density $\rho_{\phi}$ remains constant afterwards 
\footnote{This behavior 
does not occur in the situation in which the mass varies linearly with
the field, although one can let the ratio between $\rho_{\chi}$ and $\rho_{\phi}$
remain constant for a long time, until the DE finally comes to dominate completely
\cite{hoffmanvamps}.}.
This behavior could solve the ``cosmic coincidence problem,'' that is:
why are the DE and DM energy densities similar today?  

In this article, we want to study the extent of fine tuning required
in this class of models given the observational constraints on the Hubble
constant $H_0$ and the dark energy density today $\Omega_{\phi 0}$. We also compute
the resulting age of the Universe for the allowed models.
In Sec. \ref{sec:idea} we present
the main equations of the VAMP scenario and the constraints that have
already been discussed elsewhere. In Sec. \ref{sec:finetuning}
we show that based on very robust constraints 
such models are not able to solve the cosmic coincidence problem.  Section 
\ref{sec:age} discusses the constraints imposed by the age of the
Universe in these models, showing that better model-independent 
measurements of a lower limit 
would provide a possible way to
distinguish between the noncoupled and 
coupled dark energy models. In Sec. 
\ref{sec:conclusion} we present our conclusions.

\vspace{-2mm}
\section{The idea of exponential VAMPs} \label{sec:idea}
\vspace{-2mm}

In the exponential VAMP model, the potential of 
the DE scalar field $\phi$ is given by
\begin{equation} \label{eq:potential}
V(\phi)  = V_0 \ e^{\beta \phi/ m_p} , 
\end{equation}
where $V_0$ and $\beta$ are positive constants and 
$m_p = M_p / \sqrt{8 \pi} = 2.436 \times 10^{18}$ GeV is the reduced Planck mass
in natural units, $\hbar = c = 1$. 
Dark matter is modeled by a scalar particle $\chi$ of mass 
\begin{equation} \label{eq:vampmass}
M_{\chi} = M_{\chi 0} \ e ^{- \lambda (\phi- \phi_0) /m_p} \ ,
\end{equation}
where $M_{\chi 0}$ is the current mass of the dark matter particle  (hereafter the index $0$ denotes
the present epoch, except for the potential constant $V_0$) and $\lambda$ is a
positive constant. Since dark matter must be stable, its
 number density $n_{\chi}$ obeys the equation
\begin{equation} \label{eq:number}
\dot{n}_{\chi} + 3 H n_{\chi}  = 0 \ ,
\end{equation}
where the dot denotes a derivative with respect to time, $H = \dot{a}/a $, and $a(t)$ is the scale factor.
Assuming that the dark matter is nonrelativistic, its energy density is given
by $\rho_{\chi} = M_{\chi} n_{\chi}$, and it follows that
\begin{equation} \label{eq:vampfluid}
\dot{\rho}_{\chi} + 3 H \rho_{\chi} = -\frac{\lambda \dot{\phi}}{m_p} \ \rho_{\chi} \ .
\end{equation}
Since the total energy-momentum tensor has to 
be conserved, the fluid equation for dark energy is given by 
\begin{equation} \label{eq:defluid}
\dot{\rho}_{\phi} + 3 H \rho_{\phi} (1 + \omega_{\phi}) = \frac{\lambda \dot{\phi}}{m_p} \ \rho_{\chi} \ ,
\end{equation}
where  $\omega_{\phi} = p_\phi/\rho_\phi =
(\frac{1}{2} \dot{\phi}^2 - V)/(\frac{1}{2} \dot{\phi}^2 + V) $ is 
the usual equation of state parameter for an homogeneous scalar field.

These equations can also be written in the form
\begin{eqnarray} \label{eq:bothfluids}
\dot{\rho}_{\chi} + 3 H \rho_{\chi} (1 + \omega^{(e)}_{\chi}) = 0 \ , \\
\dot{\rho}_{\phi} + 3 H \rho_{\phi} (1 + \omega^{(e)}_{\phi}) = 0 \ ,
\end{eqnarray}
where
\begin{equation} \label{eq:vampeffstate}
\omega^{(e)}_{\chi} =\frac{\lambda \dot{\phi}}{3 H m_p} = \frac{\lambda \phi'}{3 m_p}\ ,
\end{equation}
\begin{equation} \label{eq:deeffstate}
\omega^{(e)}_{\phi} = \omega_{\phi} -\frac{\lambda \dot{\phi}}{3 H m_p} \ \frac{\rho_{\chi}}{\rho_{\phi}}
=\omega_{\phi} -\frac{\lambda \phi'}{3 m_p} \ \frac{\rho_{\chi}}{\rho_{\phi}}  \ 
\end{equation}
are the effective
equation of state parameters for dark matter and dark energy, respectively. 
Primes denote derivatives with respect to $u=  \ln (a) = - \ln (1+z)$,
where $z$ is the redshift, and $a_0 = 1$.
From the above equations
one can also obtain the DE equation of motion,
\begin{equation} 
\ddot{\phi} + 3 H \dot{\phi} = \frac{\lambda \rho_{\chi}}{m_p} - \frac{\beta V}{m_p} \ .
\end{equation}

The Friedmann equation for a Universe with DE, DM, baryons, and radiation is given by
\begin{equation} 
H^2 = \frac{1}{3m_p^2} \ \left[ \rho_b + \rho_r + \rho_{\chi} + \frac{1}{2} \dot{\phi}^2 + V(\phi) \right] \ , 
\end{equation} 
where we have assumed that the Universe is flat,
\begin{equation}
\Omega_0 \equiv \frac{\rho_0}{\rho_{c0}} = \Omega_{\phi 0} + \Omega_{\chi 0} + \Omega_{b0} + \Omega_{r0} = 1 \ ,
\end{equation}
where $\rho_{c0} = 3 m_p^2 H_0^2 = 8.1\; h^2 \times 10^{-47}$ GeV$^4$ is the present critical density 
and the Hubble parameter is $H_0 = 100 h$ km s$^{-1}$ Mpc$^{-1}$.
$\rho_b$ and $\rho_r$ denote the energy densities of baryons and radiation, respectively, which
for $i =r,b$ satisfy the equations $\dot{\rho}_i + 3 H \rho_i (1 + \omega_i)=0$, where $\omega_b = 0$
and $\omega_r = 1 / 3$. 
It is useful to express $V_0$ in units of 
\begin{equation} \label{eq:v0unit}
\widetilde{\rho_c} = 4.2 \times 10^{-47} \ \mathrm{GeV}^4 \ , 
\end{equation}
the critical density value for $h=0.72$.
Observational limits on the Hubble parameter gives $h=0.72 \pm 0.08$ \cite{freedman}, 
which we adopt in what follows.

The  Friedmann and DE motion equations become, in terms of $u$,
\begin{eqnarray} \label{eq:demotion}
H^2 \phi'' & + & \frac{1}{3 m_p^2} \left[\frac{3}{2} ( \rho_b + \rho_{\chi}) + \rho_r + 3 V \right] \phi' 
\nonumber \\ [-2mm]
& & \\ [-2mm]
& = & 
\frac{\lambda \rho_{\chi}}{m_p} - \frac{\beta V}{m_p} \ ,  \nonumber
\end{eqnarray}
\begin{equation} \label{eq:friedmann}
H^2 = \frac{(1/3 m_p^2) \ (\rho_{\chi} +\rho_b + \rho_r + V )}{1 - (1/6 m_p^2) \ {\phi'}^2} \ .
\end{equation}
%
%
%
%
Using the fact that the right-hand side of Eq. (\ref{eq:demotion}) 
is the derivative with respect to the field $\phi$ 
of an effective potential \cite{hoffmanvamps},
\begin{equation} \label{eq:effpotential} 
V_{eff}(\phi) = V(\phi) + \rho_{\chi}(\phi) \ ,
\end{equation}
one can show that there is a fixed point value for the field, given by $d
V_{eff}(\phi)/d \phi=0$:
\begin{equation} \label{eq:field}
\frac{\phi}{m_p} = - \ \frac{3}{(\lambda+\beta)} \ u \ + \  \frac{1}{(\lambda+\beta)} 
\ln \left( \frac{\beta V_0}{\lambda \rho_{\chi 0} e^{\lambda \phi_0/m_p}} \right) \ .
\end{equation}
At the present epoch the energy density of the Universe is divided 
essentially between
dark energy and dark matter. In this limit, using the above solution, 
one obtains
\begin{equation} \label{eq:fpOmega}
\Omega_{\phi} = 1 - \Omega_{\chi} = \frac{3}{(\lambda + \beta)^ 2} + 
\frac{\lambda}{\lambda + \beta} \ ,
\end{equation}
\begin{equation} \label{eq:fpomega}
\omega^{(e)}_{\chi} = \omega^{(e)}_{\phi} = - \ \frac{\lambda}{\lambda + \beta} \ ,
\end{equation}
which is a stable attractor for  $\beta > -\lambda/2 + (\sqrt{\lambda^ 2 + 12}) / 2$ 
\cite{exp1amendola,riottovamps}. The equality between $\omega_{\chi}$ 
and $\omega_{\phi}$ in the attractor regime
comes from the tracker behavior of the exponential potential 
\cite{exp1amendola,riottovamps,exponentialscaling} in this regime. 

\begin{figure}
\includegraphics[scale=0.35,angle=90]{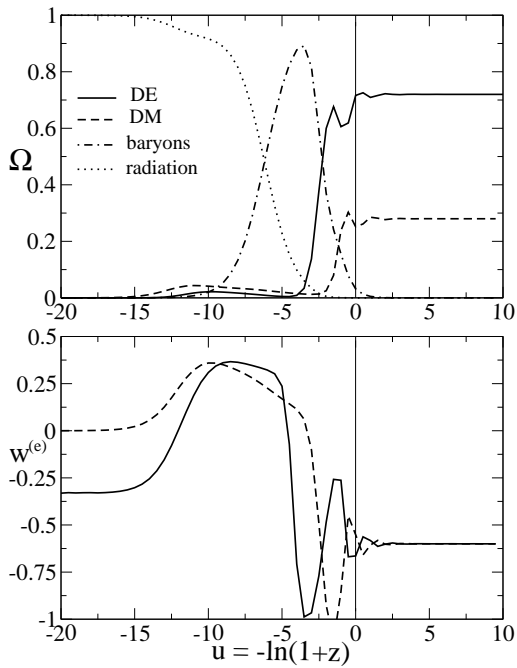}
\caption{\label{fig:omegas} \small {\it Top panel}: Density parameters of the 
components of the Universe
as a function of $u=- \ln(1+z)$ for $\lambda = 3$, $\beta = 2$, and 
$V_0 = 0.1 \widetilde{\rho_c}$. 
After a transient period of baryonic matter domination (dot-dashed line), DE comes
to dominate and the ratio between the DE (solid line) and DM (dashed line) energy densities 
remains constant.
{\it Bottom panel}: Effective equations of state for DE (solid line) and DM (dashed line)
for the same parameters used in top panel. In the tracker regime both equations of state 
are negative.}
\end{figure}

The density parameters for the components 
of the Universe and the effective equations of state for the 
DE and DM for a typical solution are shown in Fig. \ref{fig:omegas}.
Notice that the transition to the tracker behavior in this example is 
currently occurring. 

Let us assume
observational upper and lower limits to the DE equation of state parameters  
$ \omega_{obs}^{l} \leq \omega^{(e)}_{\phi 0} \leq \omega_{obs}^{u}$ from SNIa, 
for instance. Then, from Eq. (\ref{eq:fpomega}) it  
follows a constraint on the parameter space ($\lambda,\beta$) given by
\begin{equation} 
\beta \leq (\geq) \ \frac{1 - |\omega_{obs}^{u(l)}|}{|\omega_{obs}^{u(l)}|} 
\ \lambda \ \ \  ,
\label{eq:wconstraint}
\end{equation}
which is valid if the field has already reached the attractor solution. 

%
%
%

Analogously, in the attractor solution, the density parameter $\Omega_{\phi 0}$ 
can also constrain the parameter space, since Eq. (\ref{eq:fpOmega})
implies
\begin{eqnarray} \label{eq:Omegacostraint}
\beta & \geq &(\leq) \  \frac{(1 - 2 \ \Omega_{obs}^{(DE)u(l)})}{2 \ 
\Omega_{obs}^{(DE)u(l)}} \ 
\lambda  \\ 
& & \nonumber \\ [-1mm]
& + & 
\frac{\sqrt{\lambda^2 [{(2 \ \Omega_{obs}^{(DE)u(l)} - 1)}^2 - 
4 (\Omega_{obs}^{(DE)u(l)} -1)] + 12}}{2 \ \Omega_{obs}^{(DE)u(l)}} \ , 
\nonumber 
\end{eqnarray}
where $\Omega_{obs}^{(DE)u(l)}$ is the upper (lower) limit to the 
current value of DE density parameter. 

\begin{figure}
\includegraphics[scale=0.35]{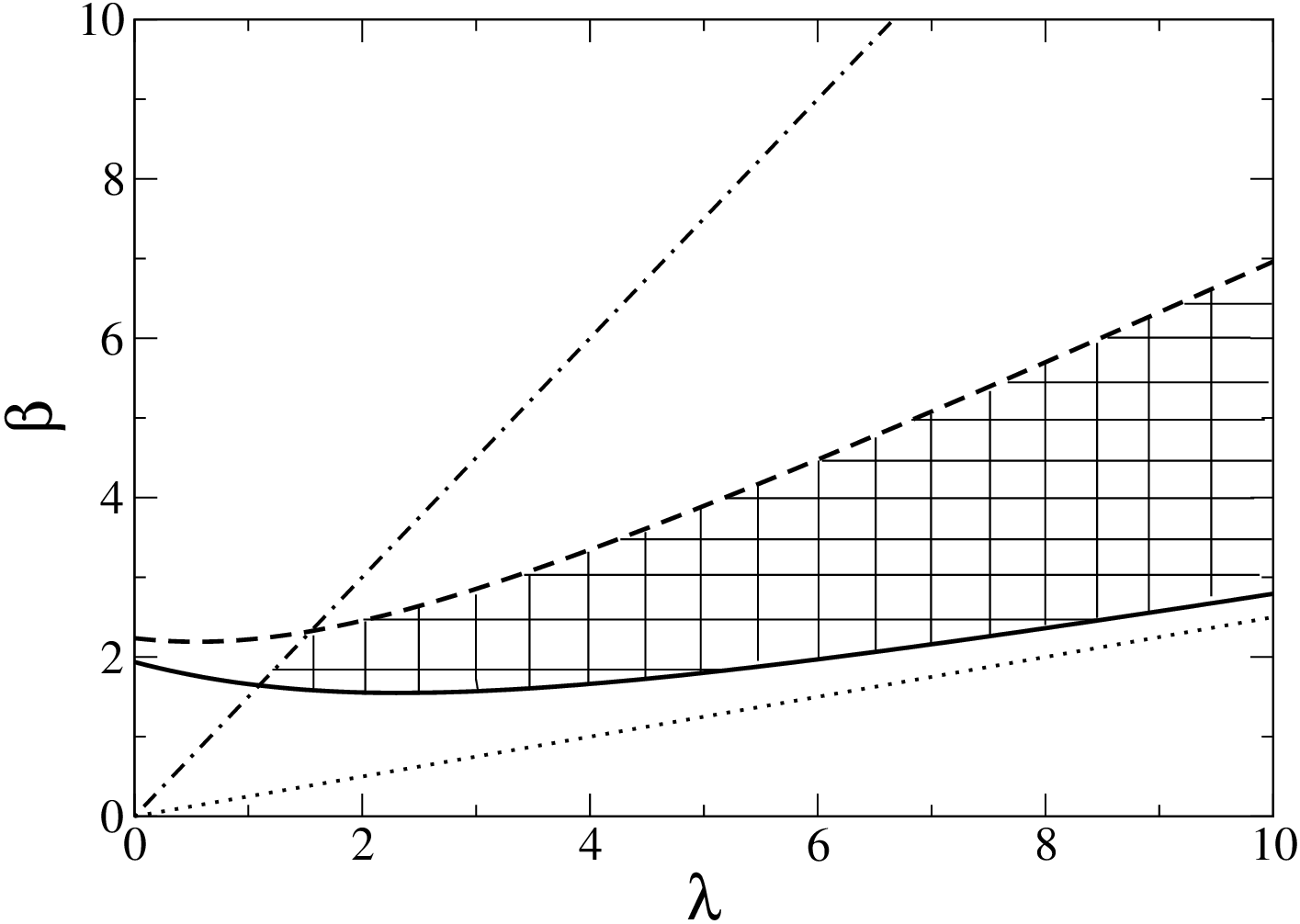}
\caption{\label{fig:paramspace} {\small Constraints on parameter space 
($\lambda,\beta$) given by DE equation of state and DE density parameter. 
The upper (lower) limit on equation of state allows the regions under (above) 
the dot-dashed (dotted) 
line for $\omega_{obs}=-0.4 \ (-0.8)$.
On the other hand, the upper (lower) limit on the DE density parameter, 
$\Omega_{obs}^{(DE)} = 0.8 \ (0.6)$, 
requires the parameters to be above (under) the solid (dashed) line. The
hatched region is allowed by both constraints.}}
\end{figure}

Amendola \cite{exp3amendola} has studied in detail the observational 
constraints imposed in this scenario by the dark energy equation of state 
(obtained from luminosity distance consistent with high $z$ 
supernovae) and density parameters when the attractor regime 
has already been reached. He obtained $-0.8 \leq \omega_{\phi 0}^{(e)} \leq -0.4$
at 95$\%$ confidence limit (C.L.). Figure \ref{fig:paramspace} shows
the allowed region in the ($\lambda,\beta$) parameter space using 
this result together with a conservative
bound $0.6 \leq \Omega_{\phi 0}\leq 0.8$ on the DE density parameter. 

%

\section{Fine Tuning} \label{sec:finetuning}

Tracker solutions of noncoupled exponential quintessence models in which 
the dark energy equation of state tracks that of the background are not 
able to accelerate the Universe today. 
Exponential VAMPs solve this problem because the background acquires 
a negative equation of state. However, it is necessary to verify 
whether fine tuning of the model is necessary to explain the observations today.

In order to quantify the amount of 
fine tuning required, we 
have solved numerically the coupled equations of motion for the  
scalar field, varying $V_0$, $\lambda$, and $\beta$ in the region $\lambda =
[0.01,20]$, $
 \beta= [0.01,20]$, and $V_0= [0.01 \widetilde{\rho_c}, 1.5\widetilde{\rho_c}]$,
 with step sizes 
$\Delta \lambda = \Delta \beta = 0.25$ and 
$\Delta V_0 = 0.05 \widetilde{\rho_c}$, generating about $1.9 \times 10^5$ models,
from which we have selected those that agree with observed 
cosmological parameters. 

\begin{figure}
\includegraphics[scale=0.35]{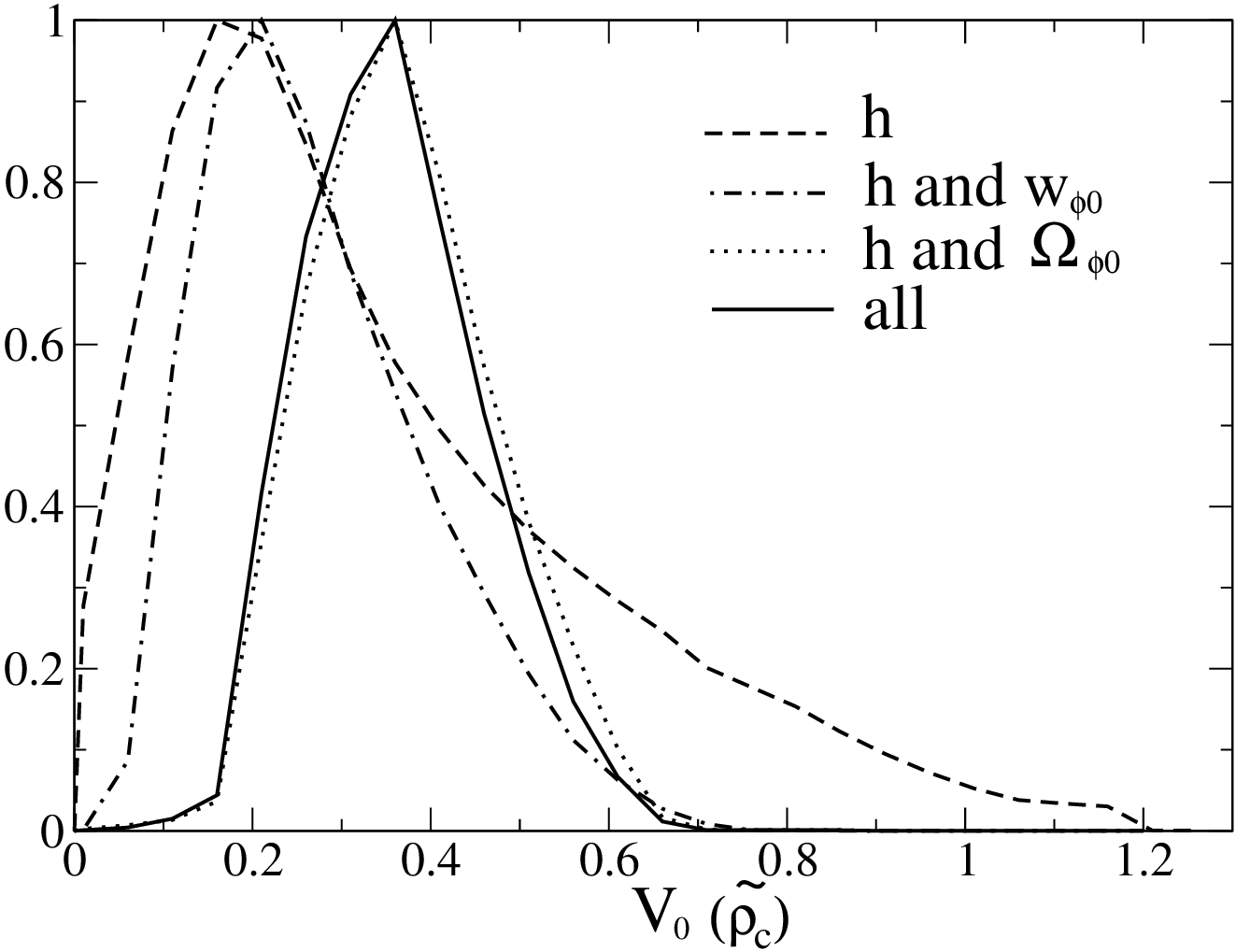}
\caption{\label{fig:v0} {\small Normalized number of models that satisfy 
the observational constraints discussed in the text.}}
\end{figure}

Figure \ref{fig:v0} shows the normalized number of models 
that satisfy some observational constraints as function of
$V_0$, given in units of $\widetilde{\rho_c}$. 
The dashed line shows the normalized number of models 
that satisfy the Hubble parameter constraint, $h=0.72\pm 0.08$ \cite{freedman}.
Notice that in order for this constraint to be satisfied it
requires $V_0$ to be of the order of $\widetilde{\rho_c}$, the critical density 
today.
The dot-dashed line is obtained requiring both the 
Hubble parameter constraint and the equation of state constraint,
$-0.8 \leq \omega_{\phi 0}^{(e)} \leq-0.4$ at 95$\%$ C.L. \cite{exp3amendola}.
The dotted line corresponds to the case when the Hubble parameter constraint
and parameter density constraint, $0.6 \leq \Omega_{\phi 0} \leq 0.8$, 
are taked into account, and the solid line when these three
parameters are constrained simultaneously. One can see that the narrow allowed 
range of $V_0 = [0.25\widetilde{\rho_c},0.45\widetilde{\rho_c}]$, 68$\%$ C.L.
($V_0=[0.15\widetilde{\rho_c},0.55\widetilde{\rho_c}]$, 95$\%$ C.L.),
is essentially determined by the constraints given by $h$ and 
$\Omega_{\phi 0}$, which are very conservative and robust. Hence, fine tuning in
this class of models seems unavoidable.

%
%
What does this fine tuning in $V_0$ mean?  
$V_0$ is the main parameter that determines the epoch in which
the field reaches the attractor regime. 
As can be seen in Fig.
\ref{fig:omegas}, for the allowed values of $V_0$, the Universe is
entering the attractor regime around the present epoch. This implies that 
the equation
of state and the parameter density are still varying today, although 
their current values are very similar to the values at the fixed point.

%
\begin{figure}
\vspace{0.5cm}
\includegraphics[scale=0.35]{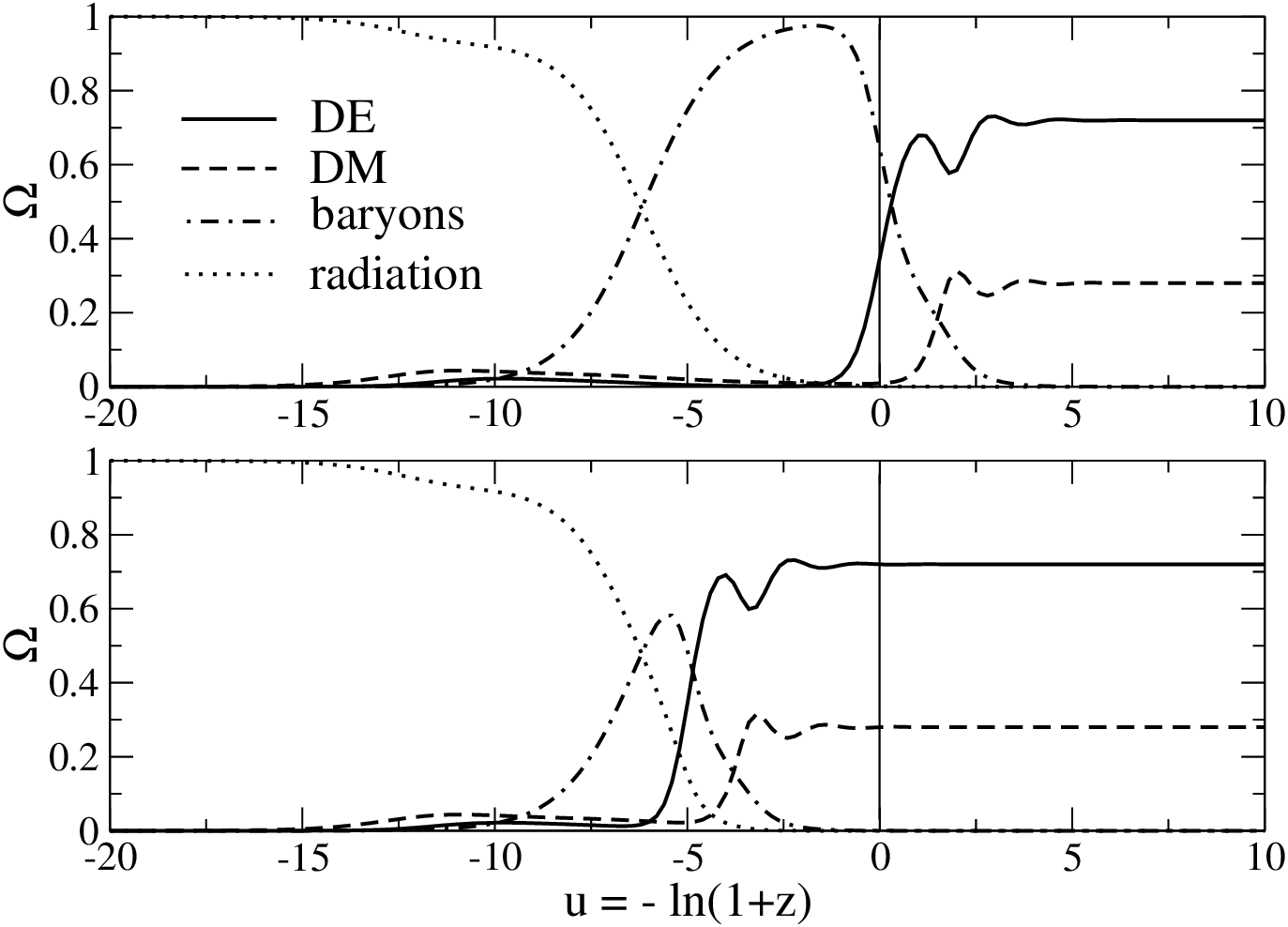}
\caption{\label{fig:attractor} {\small Density parameters of the components
of the Universe as function of $u=\ln(1+z)$ for $\lambda=3$, $\beta=2$. In the
top panel $V_0 = 10^{-4} \widetilde{\rho_c}$, and in the bottom panel 
$V_0 = 10^{3} \widetilde{\rho_c}$.
}}
\end{figure}
%
However, if one varies $V_0$ a similar behavior of $\Omega_{DE}$ and 
$\Omega_{DM}$ sets in earlier or later, as can be seen in  
Fig. \ref{fig:attractor}.
However, unrealistic values of the Hubble parameter result in this case. For
instance, for $V_0 = 10^{3} \widetilde{\rho_c}$, where the fixed point is
reached earlier, we obtain $h \approx 10$. In this sense, the cosmic 
coincidence problem is not solved in these models,
since one has to choose the overall scale of the 
potential in order to obtain realistic cosmological
solutions, very much like noncoupled dark energy models.  

\section{Age of the Universe} \label{sec:age}

The potential to use the age of the Universe as a constraint 
to dark energy models has been discussed in several 
recent papers \cite{kraussscience,krauss,jimenez,copeland}.
The WMAP collaboration \cite{wmap} has given a stringent value 
for the age of the Universe,
$t_0 = 13.7 \pm 0.2$ Gyr at 68$\%$ C.L.
However, this result is model dependent and obtained from
direct integration of the Friedmann equation for the
running spectral index $\Lambda$CDM model (cold dark matter
model with a cosmological constant).
Observations also
indicate a lower limit for the age of the
oldest globular clusters (and consequently for the age of the Universe)
of $t_0 = 10.4$ Gyr  at 95$\%$ C.L. \cite{kraussscience}. 

Using the limits discussed above and 
the position of the first Doppler 
peak in the WMAP experiment it is possible to put an upper 
limit on the equation of state (assumed constant) of 
the dark energy, $\omega_{\phi} < -0.67$ at 90$\%$ C.L. 
\cite{jimenez}. However, current upper limits on the 
age of the Universe alone cannot constrain the lower value
of the equation of state, since for large values
of the age ($t_0 \gtrsim 18$ Gyr) the cosmology is 
essentially independent 
of the value of the equation of state \cite{krauss},
unless we assume that reionization took place too early
in the Universe history (as indicated by WMAP data), 
and consequently the upper
limit for the age of the Universe is close to the upper
limits coming from globular clusters, $t_0 \approx 16$ Gyr,
as in the case discussed in Ref. \cite{jimenez}.

A model independent approach to estimating the age of the 
Universe has been performed by means of a parametrization for
the equation of state of dark energy that allows 
it to vary \cite{copeland}. Using cosmic microwave background 
(CMB) and supernova data
the age of the Universe was found to be 
$t_0 = 13.8 \pm 0.3$ Gyr at 68$\%$ C. L.
Even this model independent approach, however, is not able 
to parametrize the VAMP scenario, since in this case
the effective equation of state of the CDM is also
variable, and the strong limits in the 
age of the Universe obtained by these CMB analyses 
cannot be used to constrain the
models discussed here. 
For this reason, 
we have used only the conservative limits obtained from
globular clusters to study the potential
of the age of the Universe to constrain 
this scenario. Therefore we adopt
$10.7 < t_0 <16.3$ Gyr at 95$\%$ C.L.
\cite{kraussscience}. We have added to the age of globular clusters 
0.3 Gyr, since the WMAP data indicates an early reionization
of the Universe \cite{wmap,jimenez}. The age of the Universe 
in the VAMP scenario is given by
\begin{eqnarray} \label{eq:vampst0}
t_0  & = & H_0^{-1}
\int_{0}^{z_i} \left\{ \Omega_{\phi 0} \exp\left[3 \int_{-u_i}^{0} [
1 + \omega^{(e)}_{\phi}(u)] 
du \right]  \right.
\nonumber \\ [-1mm]
\nonumber & & \\ 
& + & 
\left. \Omega_{b0} (1+z)^3 + \Omega_{r0} (1+z)^{4} \right.
 \\ [-1mm]
\nonumber & & \\ [-1mm]
& + & 
\left. \Omega_{\chi 0} \exp\left[3 \int_{-u_i}^{0} \left[1 + 
\omega^{(e)}_{\chi}(u) \right]du \right]\right\}^{-1/2}  \frac{dz}{(1+z)} \ , \nonumber
\end{eqnarray}
where $u_i$ is the initial value for $u$ ($u_i =-$30 for our 
numerical calculations) and $z_i$ its corresponding redshift.

\begin{figure}
\includegraphics[scale=0.35]{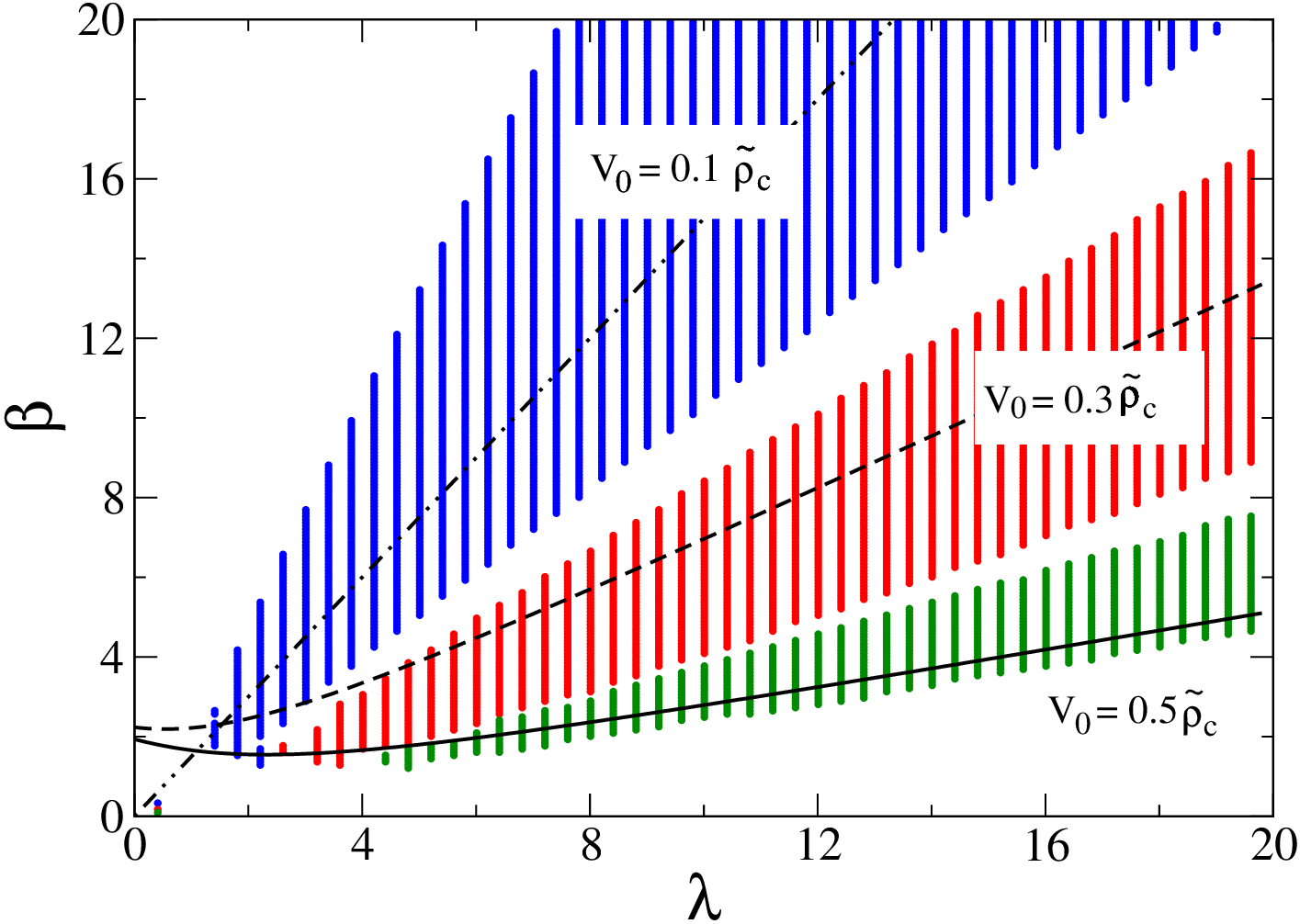}
\caption{\label{fig:paramspace2} {\small Constraints os parameter
space ($\lambda,\beta$) imposed by the age of the Universe for
$V_0 = 0.1 \widetilde{\rho_c}$, $V_0 = 0.3 \widetilde{\rho_c}$, and 
$V_0 = 0.5 \widetilde{\rho_c}$. Lines 
correspond to the same limits from Fig. \ref{fig:paramspace}}.}
\end{figure}

To explore the region of parameter space that is able to satisfy the
age constraint we have used the same procedure described 
in the preceding section, varying $\lambda$
and $\beta$ in the region $[0.01,20]$, with a 
step size $\Delta \lambda = \Delta \beta = 0.05$ and a fixed value of
$V_0$, generating about $1.6 \times 10^5$ models. Models were required to satisfy 
the constraints on the Hubble parameter and on the age 
of the Universe.
Results are shown in 
Fig. \ref{fig:paramspace2}, for three different values of $V_0$. 

\begin{figure}
\includegraphics[scale=0.35]{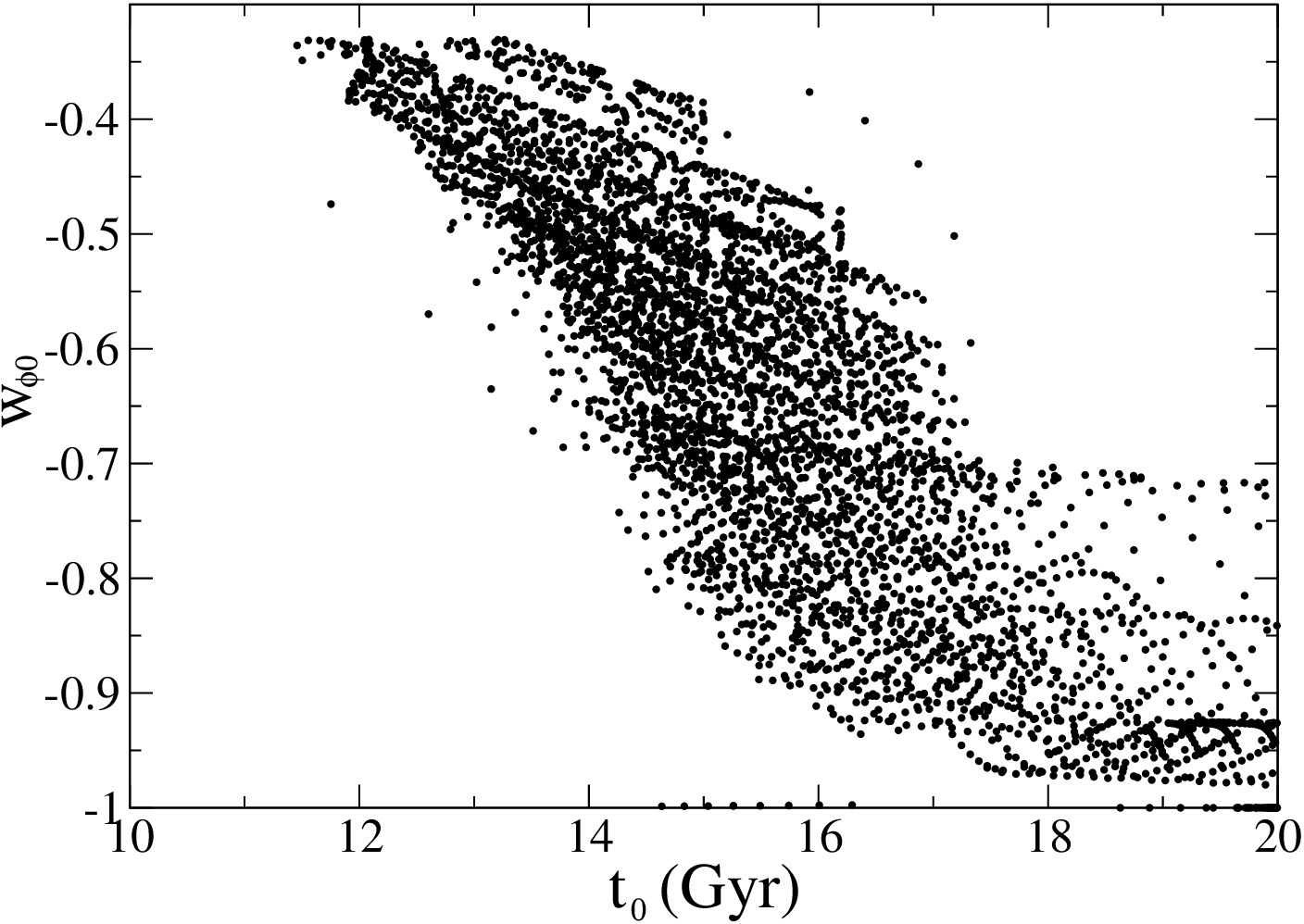}
\caption{\label{fig:t0w0} {\small Age of the Universe 
versus the present dark energy equation of state
for models that satisfy the Hubble parameter constraint.}} 
\end{figure}

The constraints imposed by the age of the Universe
are in the same direction of those imposed 
by the dark energy density and the equation of state,
indicating a correlation among them. This can be seen explicitly 
in Fig. \ref{fig:t0w0}.
Notice that such a correlation is exactly the opposite
that occurs when the equation of state (for noncoupled
dark energy) is taken to be constant \cite{jimenez}: higher values 
of $\omega_{\phi 0}^{(e)}$ imply lower ages of the
Universe. 

In the constant case, the relation between 
the age and $\omega$ is direct: for 
more negative equations of state the contribution
of dark energy becomes important later (since 
$\rho$ scales as $a^{-3(1 + \omega)}$). Consequently, for a
longer time the Universe is matter dominated, and its
age is lower compared with a Universe in 
which the dark energy has a more positive equation of state.

For the VAMP scenario, however, the situation is a little 
more complicated: models that satisfy the age of the Universe consist
of a positive correlation in the parameter space that has a 
lower slope for larger $V_0$ (Fig. \ref{fig:paramspace2}). 
For larger $V_0$, the Universe reaches the attractor 
regime earlier, and consequently has a larger value
of the age of the Universe (for the same $\beta$ and $H_0$), since
the Universe is accelerating for a longer time. Besides 
that, for the same $\beta$, larger values of $V_0$ imply
larger values of $\lambda$ for the models that satisfy
the age constraint, and consequently, from
Eq. (\ref{eq:fpomega}), the equation of state tends to $-1$.
Despite this correlation, due the 
large degeneracy in the allowed region, it seems to be
very difficult to constrain $\omega_{\phi0}^{(e)}$ based
on limits on $t_0$ and vice versa.

%
%

\begin{figure}
\vspace{0.5cm}
\includegraphics[scale=0.35]{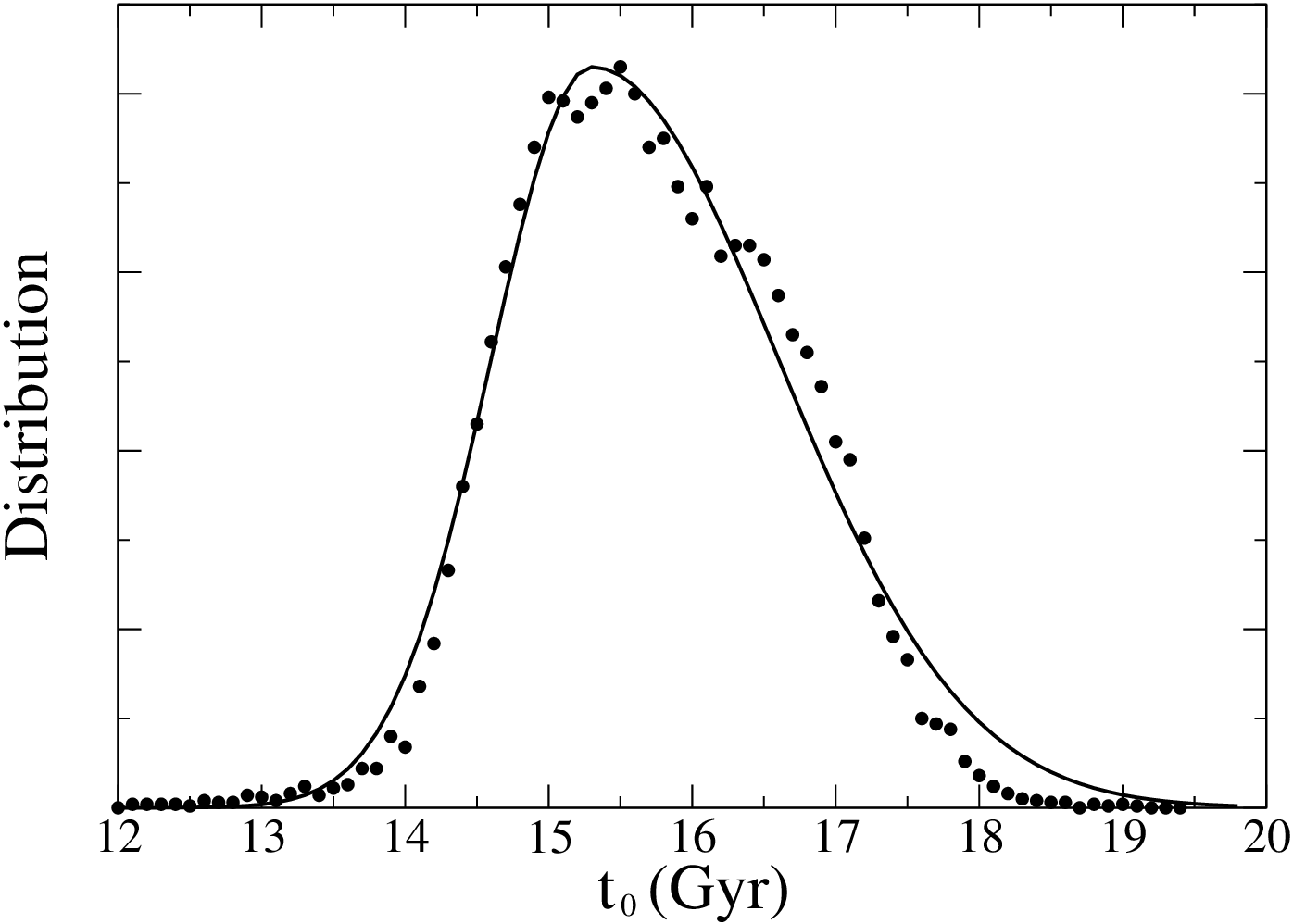}
\caption{\label{fig:age} {\small Distribution of models that satisfy 
Hubble and DE density parameters constraints 
as a function of age of the Universe. Fit to the points gives 
$t_0 = 15.3^{+1.3}_{-0.7}$ Gyr at 68$\%$ C.L.
}} 
\end{figure}

Figure \ref{fig:age} shows the age of the Universe for 
the models that satisfy the Hubble parameter and the 
dark energy density observational constraints. Here we have used
step sizes $\Delta \lambda = \Delta \beta = 0.2$ and $\Delta V_0 = 0.05 
\widetilde{\rho_c}$
for the region $\lambda = [0.01,20]$, $\beta = [0.01,20]$, 
$V_0 = [0.1\widetilde{\rho_c}, 0.8\widetilde{\rho_c}]$.

Fitting the distribution of models as function of ages, the age of the
Universe in this VAMP scenario was found to be  
\begin{equation} \label{eq:vampsage}
t_0 = 15.3 ^{+1.3}_{-0.7} \ \mathrm{Gyr} \ \mathrm{at} \ 68 \% \  \mathrm{C.L.} \ , 
\end{equation}
which is considerably higher than the age of models 
of noncoupled dark energy \cite{wmap,copeland}. 
This seems natural, since in these models
the CDM also has an effective negative equation of state and accelerates
the Universe. Thus, measurements of the age  of the 
Universe could help to distinguish between coupled and 
noncoupled models. This result is very conservative, since it relies 
only on the well established limits of the Hubble constant and the dark energy
density today. 

%
%
%

\begin{figure}
\vspace{0.5cm}
\includegraphics[scale=0.35]{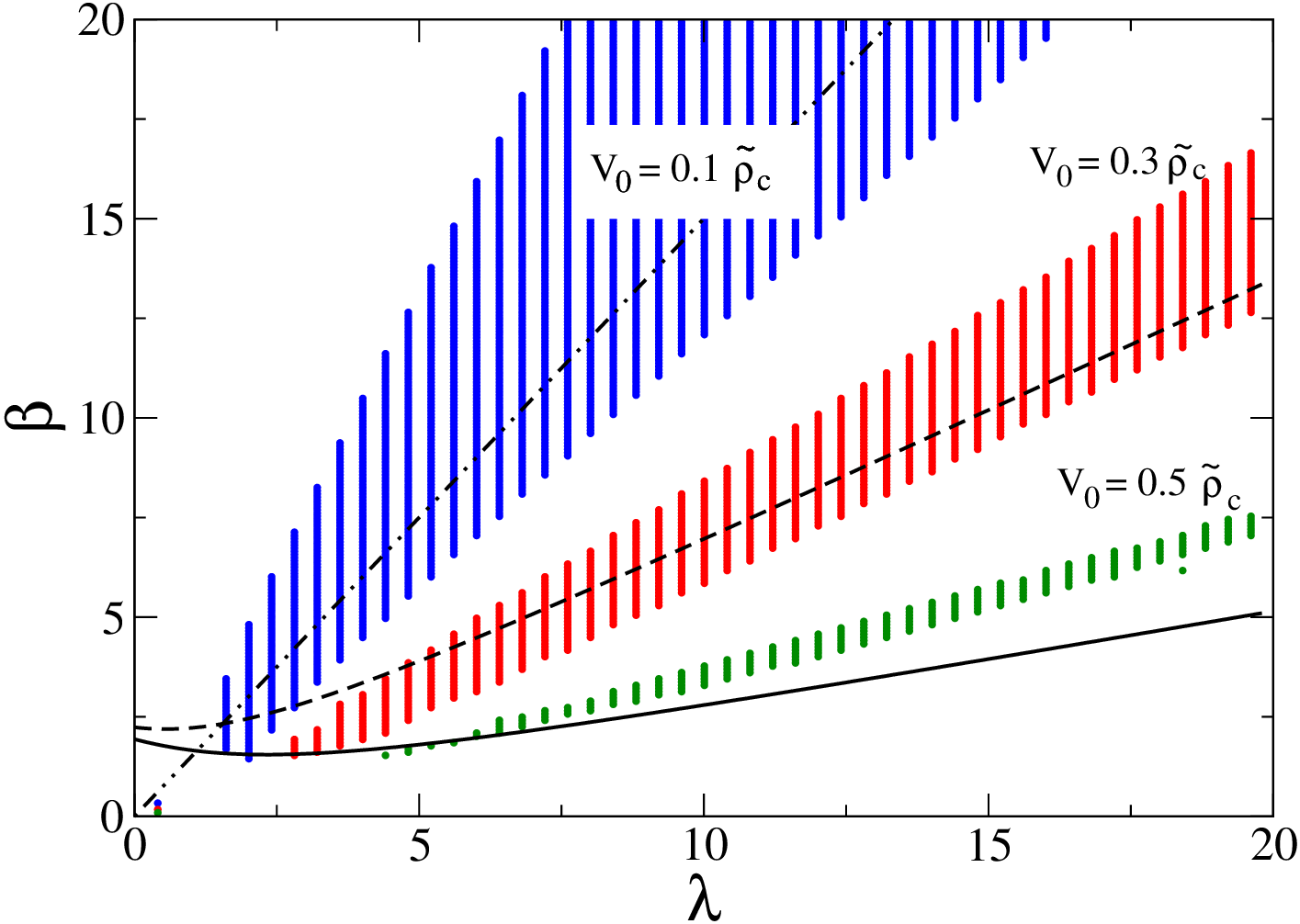}
\caption{\label{fig:paramspace3} {\small Constraints os parameter
space ($\lambda,\beta$) imposed by $t_0 = 12.5 \pm 2.2$ Gyr, and adding
$0.3$ Gyr as the age of the Universe in the epoch the globular 
cluster were formed,
for
$V_0 = 0.1 \widetilde{\rho_c}$, $V_0 = 0.3 \widetilde{\rho_c}$, and 
$V_0 = 0.5 \widetilde{\rho_c}$. Lines 
correspond to the same limits from Fig. \ref{fig:paramspace}}.}
\end{figure}

One can see that lower limits on the age of the Universe are not useful to
constraint these models.
Upper limits on the age, however, are potentially interesting.
As an example, we can 
speculate what would change if the limits on the age of 
globular clusters \cite{kraussscience,jimenez} 
were symmetric: $t_0 = 12.5 \pm 2.2$ Gyr,
and add to this age $0.3$ Gyr for the formation  of these
objects. In this case, the upper limit on the age of the 
Universe would be 15 Gyr, which would exclude a large part
of the models allowed by the present limits. The parameter
space for these limits (Fig. \ref{fig:paramspace3}) is 
much more constrained than the previous case,  practically excluding 
$V_0 > 0.5 \widetilde{\rho_c}$.

%
\section{Conclusion} \label{sec:conclusion}

The VAMP scenario is attractive since it could solve the 
problems of exponential dark energy, giving rise to a solution of the 
cosmic coincidence problem. However, in order to obtain
solutions that can provide realistic cosmological 
parameters, the constant $V_0$ has to be extremely
fine tuned in the range $V_0 = [0.25\widetilde{\rho_c},0.45\widetilde{\rho_c}]$
at 68$\%$ C.L.. This implies that the attractor
is being reached around the present epoch. 
In this sense, the model
is not able to solve the coincidence problem.  

We have found that there is a negative correlation
between the value of the age of the Universe
today and the current equation of state of dark energy:
larger values of $\omega_{\phi0}^{(e)}$ correspond to a
lower age of the Universe.
This result is the opposite of the case
in which the equation of state of dark energy
is constant.  

A generic feature of this class of models is that the Universe is older than
noncoupled dark energy models. In fact, we found $t_0 = 15.3^{+1.3}_{-0.7}\ $ Gyr.
Better model independent determination of the age of the Universe could help to
distinguish among different contenders for explaining the origin of the dark energy. 

\section*{Acknowledgments}

We would like to thank Z. Chacko for asking the right question. 
This work was supported by Funda\c{c}\~{a}o de Amparo
\`{a} Pesquisa do Estado de S\~{a}o Paulo (FAPESP), grant 01/11392-0,
and by Conselho Nacional de Desenvolvimento Cient\'{i}fico e 
Tecnol\'{o}gico (CNPq). 

%
%

\end{document}